# Grain Boundary Development of Silicon during Directional Solidification: A Phase-Field Study

Chuanqi Zhu[1], Yuichiro Koizumi[2], Chunwen Guo[3]

[1]*Graduate School of Engineering, Osaka University, 2-1, Yamadaoka, Suita, Osaka, Japan*
[2]*Division of Materials and Manufacturing Science, Osaka University, 2-1, Yamadaoka, Suita, Osaka, Japan*
[3]*School of Materials and Engineering, Zhengzhou University, Science Avenue 100, Zhengzhou, China*

In order to control the grain structure of multi-crystalline (mc) silicon during directional solidification, the development process of grain boundaries (GBs) with respect to the temperature gradient should be understood. A phase-field model incorporated with anisotropic interface energy and anisotropic attachment kinetic coefficient has produced the faceted shape of a growing silicon crystal, which is in agreement with experimental observation. The growth of coupled silicon grains under various growth velocities has been simulated to see the morphology of the solid-liquid front and the development process of the GBs. It has been found that the direction of GB is governed by either the kinetic rule or the equilibrium rule at the grain groove, depending on the growth velocity and the orientation relationship between grains on two sides. The GB beneath a groove with facet-facet surfaces follows the bisector of the two surfaces, while the direction of a GB stays far from the bisector when the groove has a rough surface. This research provides a numerical approach to predicting grain boundary development and gaining insights from grain structure evolution in mc-silicon, which can be potentially applied for high-efficiency and low-cost solar cells.

## I. INTRODUCTION

Understanding how the grain structure of multi-crystalline silicon (mc-Si) evolves during directional solidification is important for improving the quality of the silicon ingots used for solar cells [1]. The grain structure in mc-Si ingots contains the distribution of grain sizes, grain orientations, grain boundaries, and other crystal defects. One of the most discussed issues is the extension direction of grain boundaries between growing grains [2, 3]. Basically, the distribution of grain size and orientation in the mc-Si ingots results from the extension direction of the grain boundaries and the interaction between them. Therefore, knowing how the grain boundaries develop during directional solidification is one of the keys to improving the processing conditions in manufacturing and producing high-quality mc-silicon ingots [4].

Most grain boundaries between two silicon grains do not follow the direction of temperature gradient in directional solidification. This is mainly due to the fact that the growth shape of silicon is faceted and the growth velocity varies sharply with the crystalline orientation. It has been observed experimentally [5] that a growing silicon crystal has an octahedral shape, which is bounded by {111} faces. Computational work has also reproduced the faceted shape of a silicon crystal [6]. It has been already known that both the interface energy and attachment kinetic coefficient of the solid-liquid interface are anisotropic, which means these two quantities vary with the normal vector of the local interface. A series of computational works [7-9] have been carried out to clarify the growth mechanisms of two coupled silicon grains by incorporating the anisotropy of the interface energy, attachment kinetic coefficient, and grain boundary energy. In their works, the growth font formed by two misoriented silicon grains has a morphology similar to the one observed in experiments [10].

In the present research, phase-field simulation of silicon grain growth during directional solidification is conducted to elucidate how the direction of their grain boundary develops under varying cooling conditions. It has been carried out in the following steps: first, the faceted growth shape of single silicon grain in isothermal condition will be reproduced and compared with experimental observation to verify the validity of the phase-field model; second, coupled grains will be put under varying temperature conditions to investigate the behavior of solid-liquid growth front and the grain boundary beneath; Third, future work will be discussed based on the current findings. This work clarifies the underlying mechanism of grain boundary development and the model can be easily extended for simulating the growth of multiple silicon grains in three dimensions. The findings are expected to provide guidance for experimenters to control the grain structure of mc-Si and improve the quality of mc-silicon ingots used for solar cells.

## II. METHOD

### 2.1 Phase-field model

The phase-field model is developed based on the multi-phase-field method [11, 12]. The following equation that can describe the evolution of multiple phases is derived from total free energy including the interface energy and bulk free energy.

$$\frac{\partial \phi_i}{\partial t} = -\frac{2}{n} \sum_{j=1}^{n} m_{ij} \left\{ \sum_{k=1}^{n} \left[ \frac{1}{2}(\epsilon_{ik}^2 - \epsilon_{jk}^2) \nabla^2 \phi_k + (w_{ik} - w_{jk})\phi_k \right] - \frac{8}{\pi} \sqrt{\phi_i \phi_j} \Delta g_{ij} \right\} \quad (1)$$

The subscript letters $i, j, k$ in the phase-field equation are used as indexes of the local phases. A pair of indexes indicates the interface or interaction between two phases. The right-hand side of the above phase-field equation is composed of the interfacial term and driving force term. The gradient coefficient $\epsilon$ and the penalty coefficient $w$ in the interfacial term can be associated with interface energy $\gamma$ and width $\delta$ of the diffuse interface by the following relations:

$$\epsilon = \sqrt{\frac{8\delta\gamma}{\pi^2}} \quad (2)$$

$$w = \frac{4\gamma}{\delta} \quad (3)$$

The difference in bulk free energy between the two local phases acts as the driving force $\Delta g$, which is proportional to the undercooling $\Delta T$ with a prefactor $\lambda$:

$$\Delta g = \lambda \Delta T \quad (4)$$

Additionally, the attachment kinetic coefficient $\beta$ can be related to the phase-field mobility $m$ by

$$m = \frac{\pi^2 \beta}{8\delta} \quad (5)$$

Considering the anisotropy of interface energy and attachment kinetic coefficient, the gradient coefficient and the phase-field mobility should be expressed as anisotropic functions of the normal vector $\vec{n}$ at the local interface, which is:

$$\vec{n} = \left(\frac{\partial \phi}{\partial x}, \frac{\partial \phi}{\partial y}, \frac{\partial \phi}{\partial z}\right) \quad (6)$$

In this work, the function which represents the cubic symmetry [13, 14] is used to give anisotropy to the gradient coefficient:

$$\epsilon_{(i,j)k} = \epsilon_0 \Big[ 1 - 3\zeta_{(i,j)k} + 4\zeta_{(i,j)k} \frac{(\partial \phi_k/\partial x)^4 + (\partial \phi_k/\partial y)^4 + (\partial \phi_k/\partial z)^4}{|\nabla \phi_k|^4} \Big] \quad (7)$$

The anisotropic phase-field mobility is expressed as:

$$m_{ij} = \begin{cases} m_s, & \alpha < 2° \\ (m_r - m_s)p(\chi) + m_s, & 2° \leq \alpha < 2° + \Delta\alpha \\ m_r, & \alpha \geq 2° + \Delta\alpha \end{cases} \quad (8)$$

$$\chi = \frac{\alpha - 2°}{\Delta\alpha}$$

$$p(\chi) = \chi^3(10 - 15\chi + 16\chi^2) \quad (9)$$

in which the phase-field mobility of smooth and rough interfaces is denoted as $m_s$ and $m_r$, respectively. $\alpha$ is the angle between the normal vector of the local interface and the normal vector of the facet orientation. When $\alpha$ is lower than 2°, the interface can be considered atomically smooth, and when $\alpha$ is larger than $2° + \Delta\alpha$, the interface can be regarded as atomically rough. The variable $\Delta\alpha$ indicates the transition from a smooth to a rough interface and is chosen to be 2.87°. The phase-field mobility between the smooth and rough interfaces is interpolated by function Eq.9.

The above phase-field model involved the anisotropy of the interface energy and attachment kinetic coefficient can describe the faceted shape of silicon crystal in the time and length scale comparable to experimental observation [15] by choosing the value of model parameters as follows: the grid length and time interval of each step are chosen to be $1.0 \times 10^{-5} m$, $1.0 \times 10^{-3} s$. The diffuse interface is resolved by 5 girds. The attachment kinetic coefficient in the <100> direction and the driving force prefactor are regulated to make the interface move at a speed comparable to the experiment data [15] and have values of $1.68 \times 10^{-9} m^4 J^{-1} s^{-1}$ and $4.15 \times 10^4 JK^{-1}m^{-3}$. The interface energy in the <100> direction is $0.5 J/m^2$ and the anisotropy strength in the cubic function (Eq.6) is 0.04. The grain boundary energy is assumed to be constant and has a value of $0.9 J/m^2$ without considering its dependence on orientation relationship and inclination. This is acceptable because the grain boundaries in the current work are near or have large-angle misorientation. Some grain boundaries with coincidence site lattice (CSL) may be important in determining the grain boundary direction, but the purpose of this work is mainly about the effect of the solid-liquid growth front on the grain boundary development. Therefore, the following results and discussions do not exclude but are complementary to the CSL theory in determining the grain boundary direction.

The key factor to the facet formation in the phase-field model is the anisotropy of the attachment kinetic coefficient. Two types of facets, including {111} and {110}, are considered in this research and the values of their attachment kinetic coefficient relative to the one of {100} rough surface are determined by decomposing the velocity of the rough surface in facet orientations (Fig. 1a), which makes sure that the transition from the rough surface to the facet surface is not abrupt. Accordingly, the attachment kinetic coefficient $\beta_{111}$ and $\beta_{110}$ are chosen to be $\beta_{100}/\sqrt{3}$ and $\beta_{100}/\sqrt{2}$ respectively. Fig. 1b shows the kinetic coefficient varies with the deviation from the orientation of facet planes.

## 2.2 Model Setup

A model in two dimensions has been used to investigate how the extension direction of grain boundaries is affected by the misorientation angle between two adjacent grains and growth velocity. The width and height of the domain (Fig. 2a) are 1.28 mm and 2.56 mm, resolved by 128 and 256 grids, respectively. 1000 simulation steps represent 1s in physical time. Initially, two grains with a misorientation angle are put on the left side of the domain and the rest of the domain is set as a liquid phase. The temperature gradient G is positive from the left to the right and the temperature at the solid/liquid interface is set slightly lower than the melting point of silicon. As the simulation starts and the temperature is reduced at a constant rate, the grains grow to the right side until a steady state

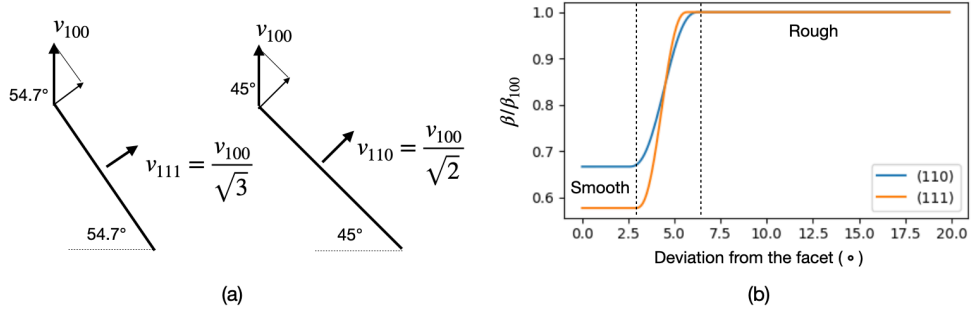

FIG. 1 (a) Decomposing the velocity of the rough surface to determine the velocity of the facet plane; (b) Attachment kinetic coefficient varies with the deviation from the surface of the facets in <110> and <111> directions.

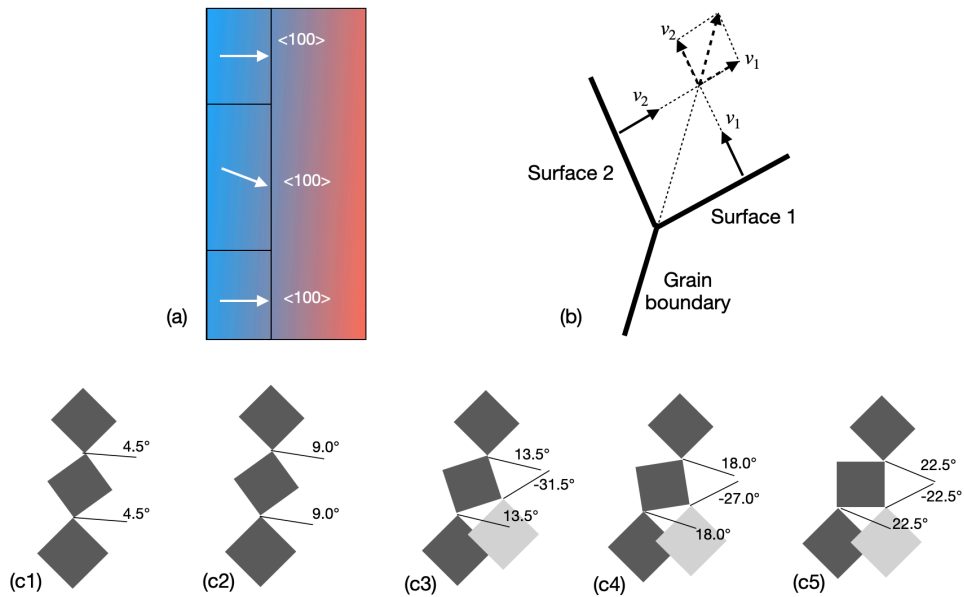

FIG. 2 (a) simulation domain with misoriented coupled grains; (b) kinetic rule for determining the grain boundary direction; (c1-5) coupled grains with increasing misorientation angles (9°, 18°, 27°, 36°, 45°) and the bisectors between faceted surfaces are denoted;

is realized. Upon the steady state, the velocity V of the solid-liquid interface is nearly equal to the value of the cooling rate $R$ divided by the temperature gradient $G$. To reduce the computation cost, the frame starts to move in the right direction and the field variables are reallocated to the computer memory when the solid-liquid interface crosses the middle of the domain width. The boundary condition of all sides is set to be symmetric.

In Fig. 2c1-5, pairs of squares with increasing misorientation angles (9°, 18°, 27°, 36°, 45°) are set up to illustrate the orientation relationships of coupled grains. Each face of the square represents the facet of a silicon crystal. The inclination of the grain boundary for a pair of facets is calculated according to the kinetic rule (Fig. 2b), which predicts the grain boundary direction by simply considering the velocities of the surfaces on the growth front. When the facets on both sides have the same velocity value, the grain boundary just follows the bisector of the angle between the facets [2]. The bisectors and their inclination angles with respect to the temperature gradient are denoted. The clockwise inclination is denoted as a positive value and the counterclockwise inclination is taken as a negative value. It should be noted that each grain boundary on the lower side in Fig. 2c3-5 has two possible ways to form a pair of facets.

The bisectors of facet pairs with acute and obtuse angles are denoted as type-A and type-B, respectively. This means the grain boundary on the lower side may choose to develop by following either bisector A or B according to the kinetic rule.

## III. RESULT

### 3.1 Equilibrium and Growth Shape of Silicon

In Fig. 3, the simulation results exhibit equilibrium and growth shapes similar to the experimental observation [5]. The initial spherical grains are put under an isothermal condition. Viewing from <110> orientation of the silicon grain, Fig. 3a and 3c show the {111} facets appear and form sharp edges under a relatively high undercooling. This crystal growth shape (CGS) forms because large undercooling gives rise to a large velocity difference between the atomically-smooth and the atomically-rough surfaces, then the edges between different surfaces become sharp. After the undercooling is decreased, shape edges become curved ones by which the surfaces are separated. This equilibrium crystal shape (ECS) results from the minimization of the interface energy rather than the velocity difference in different orientations.

The simulation also produces the same transition from CGS to ECS in the <100> orientation. The {110} facets in Fig. 3e are separated by flat and atomically-rough surfaces, which turn into curved ones in Fig. 3f when the temperature increases. The experiment and simulation results above indicate that the shape of silicon crystal in the melt is determined by both effects of the interface energy and atomic attachment, which can be regulated by the cooling condition. The similarity in crystal shape between the simulation and experiment results enables the current phase-field model to demonstrate the growth behavior of coupled grains and the development of their boundaries.

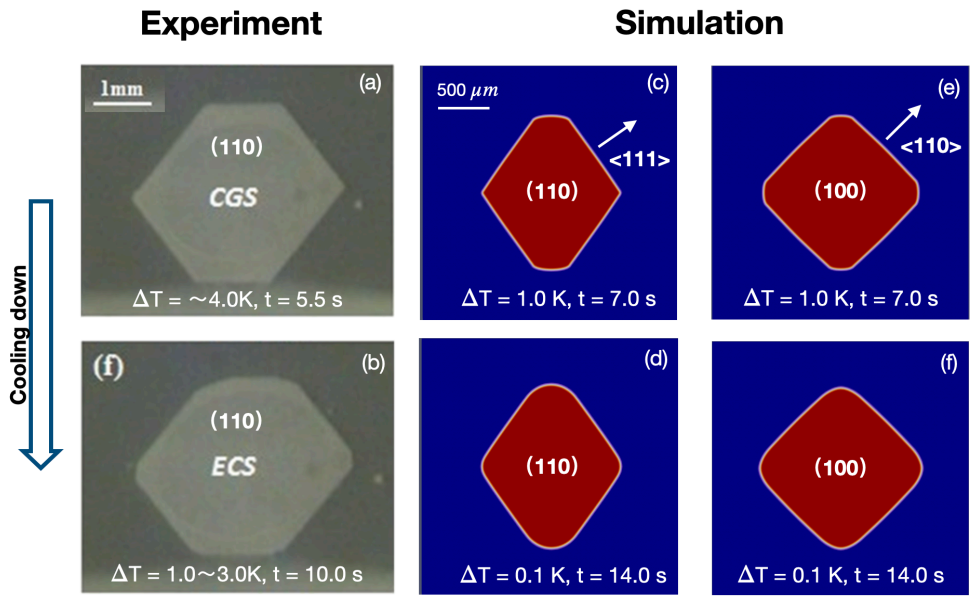

FIG. 3 Comparison between experiment [5] and simulation on the crystal growth shape (CGS) and equilibrium crystal shape (ECS) of silicon.

**3.2 Grain Boundary Development**

Fig. 4-6 show the simulation results in which the growth fronts are in steady states and grain orientation is indicated by the RGB colors. Red and green colors suggest the <100> and <110> directions are aligned with the direction of the temperature gradient, respectively. Other colors in-between suggest the increasing misorientation from left to right. The inclination angles of the grain boundaries were measured with respect to the direction of the temperature gradient. The clockwise inclination is denoted as a positive value and the counterclockwise inclination is taken as a negative value.

In Fig. 4a1-5, it is apparent that the upper grain boundaries follow closely to the bisectors denoted in Fig. 2c1-5, while the inclination of the boundaries on the lower sides is far from

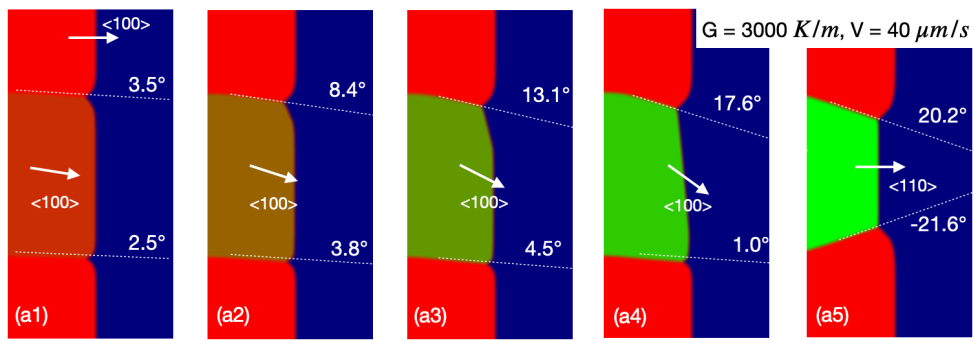

FIG. 4 (a1-5) simulation result of the growth of coupled grains in a steady-state velocity of 40 $\mu m/s$; A positive (from left to right) temperature gradient with a value of 3000 $K/m$ is the bisector inclination except for the one with 45° misorientation. This suggests that grooves of the upper GBs have the facet-facet characteristic which obeys the bisector rule, while the grooves of most lower GBs have the rough characteristic. In the simulation result of Fig. 5, for the same set of coupled grains, the growth velocity has been increased to see how the behavior of the groove and grain boundary changes. It can be perceived that at the growth velocities of 80 and 120 $\mu m/s$, the inclination angles of the upper GBs increased a little, which should be caused by the asymmetrical morphology of the groove. The direction of the lower GBs and the morphology of the grooves do not show an obvious change in comparison to the results in Fig. 4.

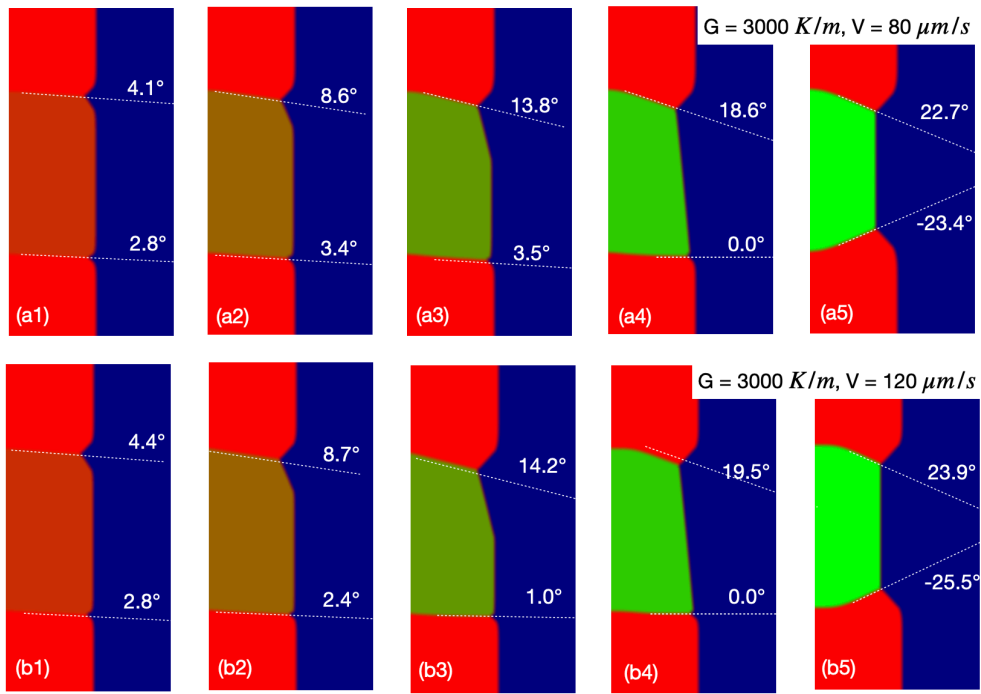

FIG. 5 Continual simulation results of steady-state growth fronts of coupled grains at growth velocities of 80 and 120 $\mu m/s$.

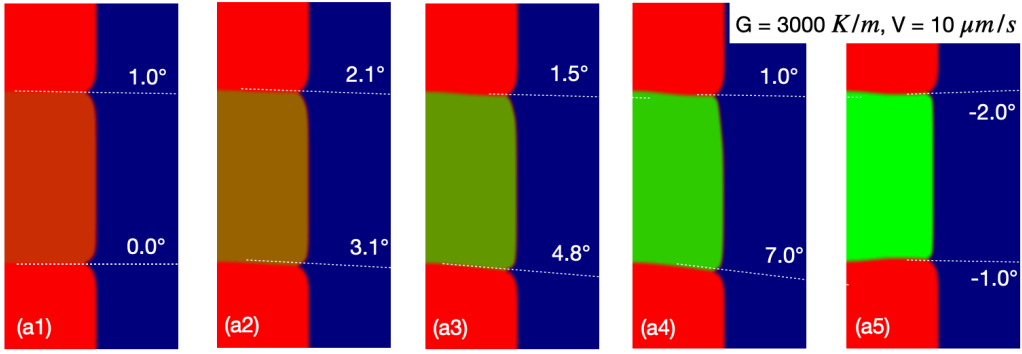

FIG. 6 Continual simulation results of steady-state growth fronts of coupled grains with a velocity of 10 $\mu m/s$.

Other than increasing the growth velocity, it is of interest to see how the coupled grains grow at low velocity. Fig. 6 shows the steady states of coupled grains at a velocity of 10 $\mu m/s$. The inclination of the upper GBs deviates abruptly from the bisectors for all misorientation angles and the one with 45° misorientation even changes to the counterclockwise inclination. The inclination of all the lower side GBs is far from that of the bisector including the one with 45° misorientation.

## IV. DISCUSSION

The plot in Fig. 7 shows how the grain boundary inclination varies with misorientation and growth velocity. The blue and orange dash lines act as the references of the type-A and type-B bisectors. The plot shows that the growth velocity has a significant effect on the inclination for both upper and lower GBs. When the growth velocity is equal to or larger than 40 $\mu m/s$, the upper GBs follow the type-A bisectors, while the lower GB is relatively close to the type-A bisector for small misorientation angle but follows the type-B bisector when the misorientation angle becomes large. When the growth velocity decreases to be as low as 10 $\mu m/s$, all the upper GBs deviate far from the type-A bisectors and the lower GB does not follow the type B bisector even for large misorientation angle. This suggests that the grooves may lose their facet-facet characteristic when growth velocity is low and the direction of GB may be governed by other rules in addition to the kinetic rule.

Fig. 8 shows the distribution map of the interface mobility extracted from the simulation results. Because the facet surface has low mobility and the rough surface has large mobility, the type of surface can be discerned in this map. All the coupled grains in Fig.8 have the misorientation of 27° and grow at an increasing steady-state velocity from the left to the right side. For the grooves of the upper GBs, the lower surfaces are facets for all growth velocities while the upper surface loses its facet characteristic when velocity is 10 $\mu m/s$. For this rough-facet groove, the GB direction is weakly governed by the kinetic rule, which states that the GB direction is aligned with the vector sum of surface velocities on two sides of the groove. It is most likely that the rule of force equilibrium at the triple junction plays a significant role in determining the upper GB direction when growth velocity is as low as 10 $\mu m/s$. The upper rough surface has larger interface energy and the lower smooth surface have smaller interface energy, the GB tends to incline counterclockwise. Although it still has a 1.5° clockwise direction, it can be reasonably expected that this GB will turn counterclockwise when further decreasing the growth velocity. Additionally, the counterclockwise upper GB in Fig. 6a5 can be well explained by the dominant effect of force equilibrium over the kinetic effect when growth velocity is low.

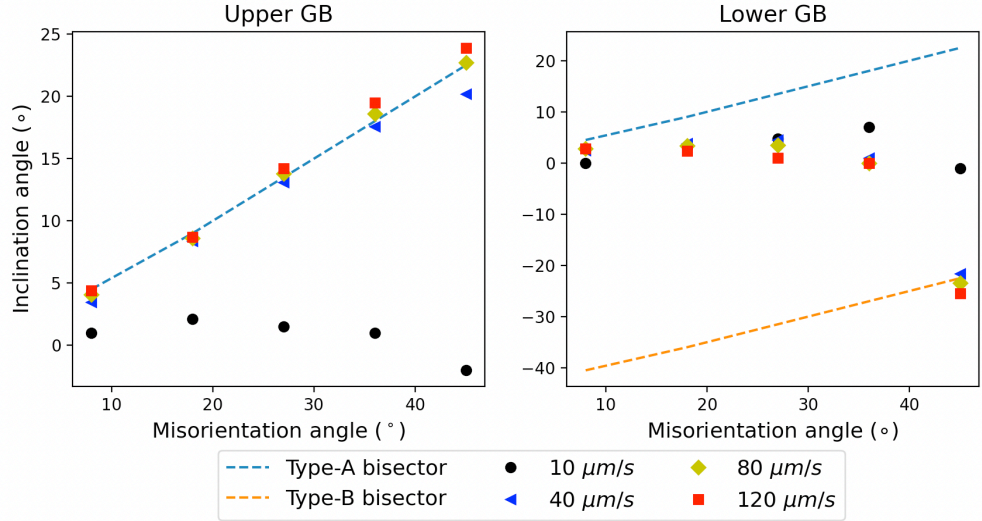

FIG. 7 Grain boundary inclination varies with misorientation angle and growth velocity.

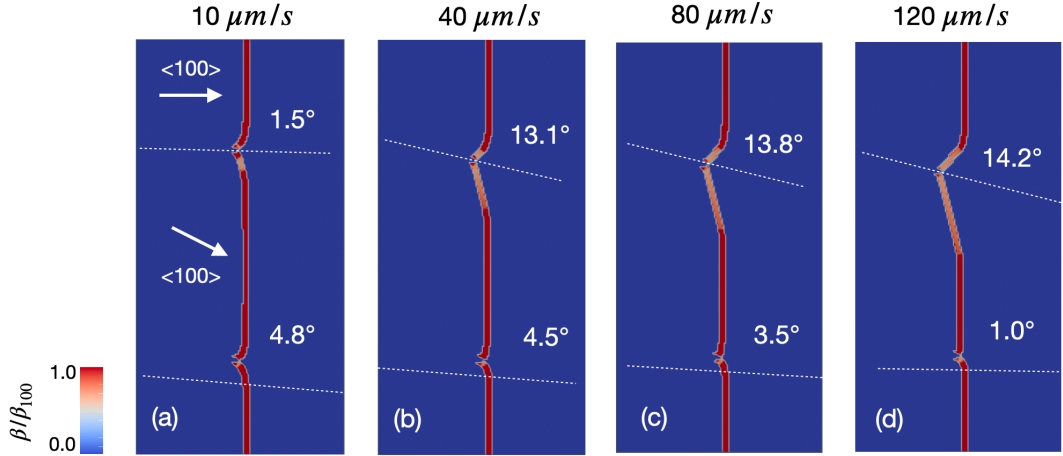

FIG. 8 The distribution map of solid-liquid interface mobility of the growth front.

It can be also discerned in Fig. 8 that for a given misorientation, while the grooves of the upper GBs become deeper with increasing velocities, the grooves of the lower GBs become more shallow and almost disappear when velocity is high. Although the lower surfaces of lower GBs tend to be facet because <100> orientation is aligned with temperature gradient, the upper rough surface with high mobility will immediately fill the gap to form a rough-rough junction with flat morphology instead of a groove. The mobility difference between the surfaces on the two sides becomes smaller when the junction is flat, and then the GB inclination decreases. According to the discussion above, the GB inclination is closely related to the type and depth of the GB groove, which is determined by both the orientation relationship between the two coupled grains and growth velocity. The following derivation explains how the groove develops under a temperature gradient.

At the steady state, the rough and smooth surfaces on the growth front move at the same velocity in the direction of the temperature gradient,

$$v = v_r = v_s \tag{10}$$

In the phase-field model, the velocity of the interface is the product of mobility and undercooling, thus,

$$v_r = m_r \Delta T_r \tag{11}$$

$$v_s = m_s \Delta T_s \tag{12}$$

The temperature difference between the smooth and rough surfaces can be expressed as,

$$\Delta T_s - \Delta T_r = \frac{v}{m_s} - \frac{v}{m_r} \tag{13}$$

The above temperature difference can also be related to the temperature gradient and displacement between smooth and rough surfaces,

$$\Delta T_s - \Delta T_r = G d \tag{14}$$

By relating Eq.11 and Eq. 12,

$$d = \frac{v}{G}(\frac{1}{m_s} - \frac{1}{m_r}) \tag{15}$$

in which the displacement $d$ is a quantity that can reflect the groove depth $D_g$. Therefore,

$$D_g \propto d \propto v(\frac{1}{m_s} - \frac{1}{m_r}) \tag{16}$$

The above expression suggests that the groove depth is not only proportional to the growth velocity but also the mobility difference between the rough and smooth surfaces, which is related to the orientation relationship between the two adjacent grains.

Fig. 9 shows the plot of groove depth varying with the growth velocity and misorientation. For the upper GB with a given misorientation, the groove depth increases with the growth velocity, while for the lower GBs except for the 45° misorientation, the groove depth is relatively unpredictable and does not show a proportionality to the growth velocity due to the rough surface in the groove. It becomes apparent that the kinetic rule is effective in predicting the GB direction when the groove has facets and large depth, while the equilibrium rule appears to become dominant in determining the inclination direction when the groove is shallow and has a rough surface.

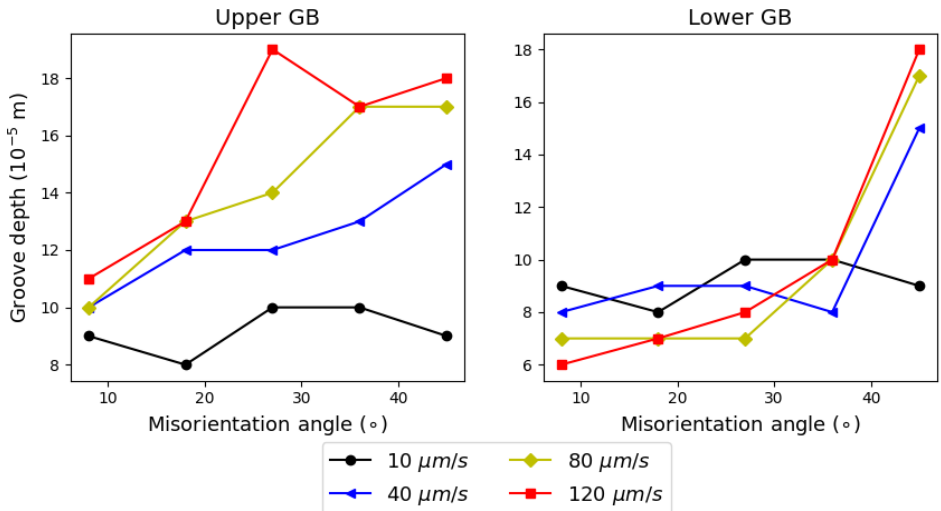

FIG. 9 Groove depth varies with increasing growth velocities and misorientation angle.

V. CONCLUSION

The development of silicon grain boundary during directional solidification has been investigated using a phase-field model and the findings are listed as follows:

- The shape of a silicon crystal responds to the undercooling due to the competitive relationship between the interface energy and attachment kinetic coefficient, which are anisotropic and have minima in the orientation of faceted and atomically-smooth surfaces.
- The grain boundary direction is closely related to the morphology and surface type of the groove, which depends on the cooling condition and orientation relationship. When the groove has a facet-facet characteristic, the GB direction follows the bisector of groove surfaces. When a non-faceted or rough surface appears in the groove, force equilibrium at the triple junction becomes dominant in determining the grain boundary direction at low growth velocity.

The above simulation is conducted for a series of misoriented coupled grains, in which the <100> orientation of one grain remains aligned with the temperature gradient. However, more possible orientation arrangements between two grains with respect to the temperature gradient can be anticipated if both of the two grains can have arbitrary orientations with respect to the temperature gradient. In addition, the orientation arrangements of grains are complex in reality because the grain growth of multi-crystalline silicon is in three-dimensional space and involved multiple grains. Therefore, in the future, the 3D simulation with multiple arbitrarily-oriented silicon grains is needed to see and understand how the grain boundaries develop and interact with each other. For this purpose, the current limited model can be improved by considering the following aspects:

First, the present model uses 2D anisotropic functions to separately describe the attachment kinetic coefficients in <100> and <110> planes. An anisotropy function of attachment kinetic coefficient in 3D is required for producing a 3D faceted silicon crystal. Second, the anisotropic function with cubic symmetry is not enough to describe the anisotropic interface energy of silicon because only small anisotropic strength can be used to maintain the stability of numerical computation. A more realistic anisotropy function for interface energy with cusps at facet orientation and a small variation in other orientations is needed. Last but not least, the grain boundary energy should depend on the orientation relationship and grain boundary inclination [16]. It may have special orientations with low energy due to coincidence site lattice (CSL), which are decisively important in determining the grain boundary direction.

Acknowledgments

This work was supported by a Kakenhi Grant-in-Aid for Scientific Research (No. 22J11558) from the Japan Society for Promotion of Science (JSPS).